\documentclass[twocolumn,showpacs, amsmath,amssymb]{revtex4}
\usepackage{graphicx}
\usepackage{dcolumn}
\usepackage{bm}

\begin{document}

\title{Scaling of impact fragmentation near the critical point}
\author{Hiroaki Katsuragi}
\email{katsurag@asem.kyushu-u.ac.jp}
\author{Daisuke Sugino}
\author{Haruo Honjo}
\affiliation{Department of Applied Science for Electronics and Materials, Interdisciplinary Graduate School of Engineering Sciences, Kyushu University, 6-1 Kasugakoen, Kasuga, Fukuoka 816-8580, Japan}
\date{\today}

\begin{abstract}
We investigated two-dimensional brittle fragmentation with a flat impact experimentally, focusing on the low impact energy region near the fragmentation-critical point. We found that the universality class of fragmentation transition disagreed with that of percolation. However, the weighted mean mass of the fragments could be scaled using the pseudo-control-parameter {\em multiplicity}. The data for highly fragmented samples included a cumulative fragment mass distribution that clearly obeyed a power-law. The exponent of this power-law was $0.5$ and it was independent of sample size. The fragment mass distributions in this regime seemed to collapse into a unified scaling function using weighted mean fragment mass scaling. We also examined the behavior of higher order moments of the fragment mass distributions, and obtained multi-scaling exponents that agreed with those of the simple biased cascade model.
\end{abstract}


\maketitle

\section{Introduction}
\label{sec:introduction}
From asteroids to nuclei, fragmentation phenomena can be seen everywhere. Many scientists and engineers have been fascinated by the fragmentation process, and much effort has been made to understand fragmentation \cite{Beysens1}. In particular, impact fragmentation of brittle solids has been investigated by simulation and experimentally \cite{Hayakawa1, Ishii1, Oddershede1, Meibom1, Kadono1}. The results have shown that cumulative fragment mass (or size) distributions exhibit a power-law dependence, in which the exponent depends on the dimensionality of the fractured objects. These results do not depend on the details of fragmentation, such as the material the fractured object is made from or the fragmentation method. Consequently, Oddershede et al. concluded that brittle fragmentation is a self-organized critical phenomenon \cite{Oddershede1,Bak1}.

Cumulative distributions obey power-law dependence when the imparted energy is sufficiently large, but brittle solids do not break when the imparted energy is very small. Therefore, one might assume that there is a transition in the fragmentation phenomenon. Recently, Kun and Herrmann examined critical behavior in fragmentation transition and concluded that fragmentation transition belongs to the percolation transition universality class \cite{Kun1}. In addition, Campi has shown that nuclei fragmentation has the same statistical properties as percolation \cite{Campi1}. By contrast, {\AA}str\"om et al.\ performed other fragmentation model simulations and obtained a scaling law whose critical exponents differ from those of percolation \cite{Astrom1}.

While there are some simulation results, there have been no experiments on critical fragmentation. For a proper analysis, the existing simulation results must be compared to experimental data. Here, we experimentally investigate brittle fragmentation as a kind of critical phenomenon, and we report experimental results for glass tube fragmentation near the critical point.

\section{Experimental procedure}
\begin{figure}
\vspace{0.25cm}
\scalebox{0.4}[0.4]{\includegraphics{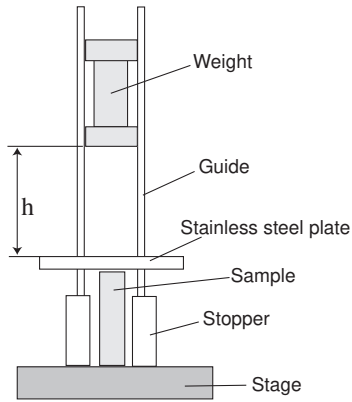}}
\caption{Schematic illustration of the experimental apparatus. The weight and guide poles were made of brass. Other parts were made of stainless steel or steel. $h$ is the falling height of the weight.}
\label{fig:apparatus}
\end{figure}

Figure \ref{fig:apparatus} is a schematic illustration of the experimental apparatus. We used glass tubes ($50$ mm outside diameter, $2$ mm thick, and $150$ mm, $100$ mm, or $50$ mm long) as the fractured objects. They have 2-D geometry, and it is easy to impart the impact vertically because they can stand by themselves. Kun and Herrmann \cite{Kun1} and {\AA}str\"om et al.\ \cite{Astrom1} performed numerical simulation of 2-D samples. 

A glass tube was put in a plastic bag, which was in turn placed between a hard stainless steel stage (dimensions $400$ mm $\times$ $500$ mm $\times$ $50$ mm) and a stainless steel plate ($10$ mm thick). The stage was fixed by placing heavy weights on it.  A cylindrical brass weight with a flat bottom ($3.77$ kg) was dropped on the stainless steel plate vertically, and a planar failure wave propagated to the glass tube from the impact circle (cross section of the glass tube). Consequently, the glass tube was cleaved by the impact.  This is a type of 2-D fragmentation. Four poles ($10$ mm diameter) were used as guide poles for the falling brass weight. Stoppers for the stainless steel plate were placed at each guide pole to prevent a secondary impact.  While the samples were not annealed after the cutting processing, they had almost the flat end. We used a level in order to check the  verticality between the weight and the stainless steel plate at each fragmentation. The variable in this experiment was the height that the weight was dropped from. 

After fragmentation, we collected the fragments in the plastic bag and measured the mass of each fragment with an electronic balance. The data were analyzed, and cumulative fragment mass distributions were obtained. We fractured a total of $60$ samples ($34: 150$ mm, $15: 100$ mm, and $11: 50$ mm length). We defined the falling height $h$ as a length from the top of stainless steel plate to the bottom of the falling weight. In our experiments, it was set in the range $61$ mm $\leq h \leq$ $407$ mm. 
This range corresponds to $0.050 \leq \epsilon \leq 0.251$ (Nm/g) in terms of the imparted energy (released potential energy of the dropped weight) per unit sample mass, $\epsilon$. 

\section{Results}
\begin{figure}
\scalebox{0.65}[0.65]{\includegraphics{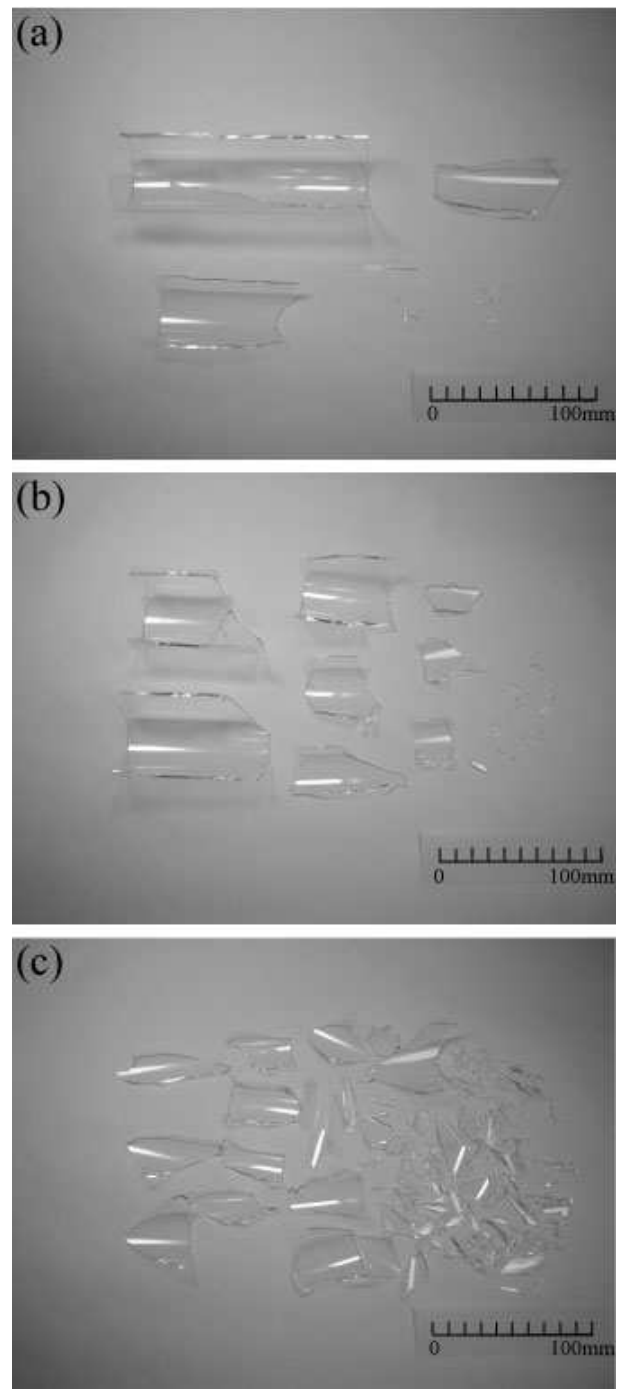}}
\caption{Typical fragments photos resulting from: (a) small imparted energy, (b) intermediate region, and (c) large imparted energy. Imparted energies per unit mass $\epsilon$ for (a), (b), and (c) are $0.098$, $0.118$, and $0.139$ (Nm/g), respectively.}
\label{fig:samples}
\end{figure}

When the height was too low to break the glass tube, no visible cracks were seen. Once a visible crack was produced, fragmentation occurred suddenly. We defined the point at which a sample began to cleave as the fragmentation critical point. We could not obtain samples that had only visible macro-cracks, but did not fragment. We show some photos of fragments in Fig.\ \ref{fig:samples}. We observed that with a flat impact, low impact energy fragmentation events produced vertical main cracks. Therefore, vertical cleaving produces a few, large fragments (Fig.\ \ref{fig:samples}(a)). In the mediate range, some large fragments remained (Fig.\ \ref{fig:samples}(b)). For large impact energy events, the largest fragment was smaller, and it was impossible to distinguish the main cracks from the many small fragments by observation (Fig.\ \ref{fig:samples}(c)). 

\subsection{Cumulative distribution}
\begin{figure}
\scalebox{1.0}[1.0]{\includegraphics{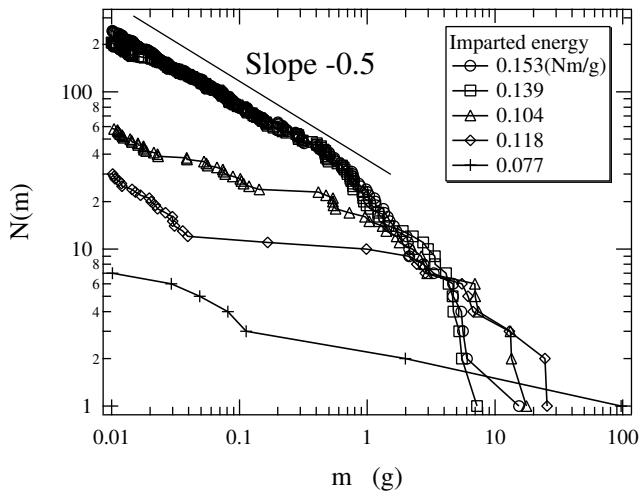}}
\caption{Some typical cumulative fragment mass distributions for $150$-mm-long samples. Power-law behavior is confirmed in the sufficiently high impact energy curves, while the distribution curves for low impact energy do not show a clear trend. The corresponding values of $\mu$ for plus, diamond, triangle, square, and circle marks are $7.05 \times 10^{-4}$, $3.32 \times 10^{-3}$, $6.47 \times 10^{-3}$, $2.34 \times 10^{-2}$, and $2.54 \times 10^{-2}$, respectively.}
\label{fig:cd}
\end{figure}

First, we plotted the cumulative mass distribution of each fragmentation. Figure \ref{fig:cd} shows some typical cumulative fragment mass distributions, $N(m)$, for $150$-mm-long samples. $N(m)$ is defined as follows:
\begin{equation}
N(m) = \int_{m}^{\infty}n(m')dm',
\label{eq:cumdef}
\end{equation}
where $m$ and $n(m)$ are the fragment mass and the number of fragments of mass $m$, respectively. The distribution curves for samples of different sizes were similar to the example in Fig.\ \ref{fig:cd}. One can see the clear power-law dependence for fully fragmented samples (far from the fragmentation transition point). The distribution curves near the critical point do not show clear power-law dependence, and they have a rather flat part. This tendency is quite different from that of percolation universality. In percolation universality, the clearest power-law dependence should appear beneath the critical point. In the fully fragmented data curves, we can observe a region that satisfies $N(m) \sim m^{-(\tau -1)}$, i.e., $n(m) \sim m^{-\tau}$. The obtained value of the characteristic exponent $\tau$ is nearly $1.5$ ($< 2$). This result conflicts with the scaling ansatz of percolation \cite{Stauffer1}. Therefore, we believe fragmentation criticality to differ from that of percolation. Conversely, our result $\tau - 1 \simeq 0.5$ concurs with {\AA}str\"om's result \cite{Astrom1}. In addition, the value of $\tau$ obtained in this experiment is consistent with Hayakawa's scaling for 2-D fragmentation with a planar failure wave \cite{Hayakawa1}.

\subsection{The weighted mean mass scaling}
\begin{figure}
\scalebox{1.0}[1.0]{\includegraphics{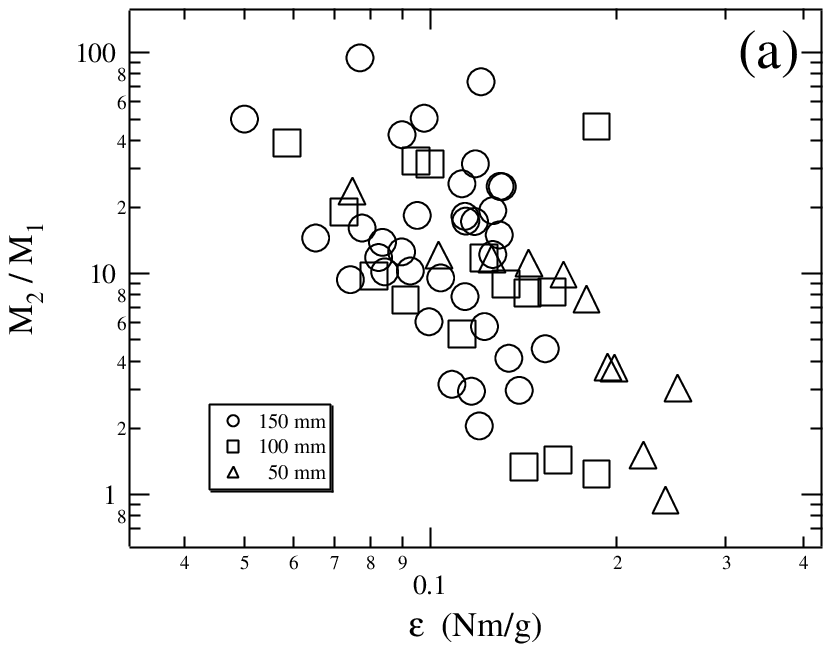}}
\scalebox{1.0}[1.0]{\includegraphics{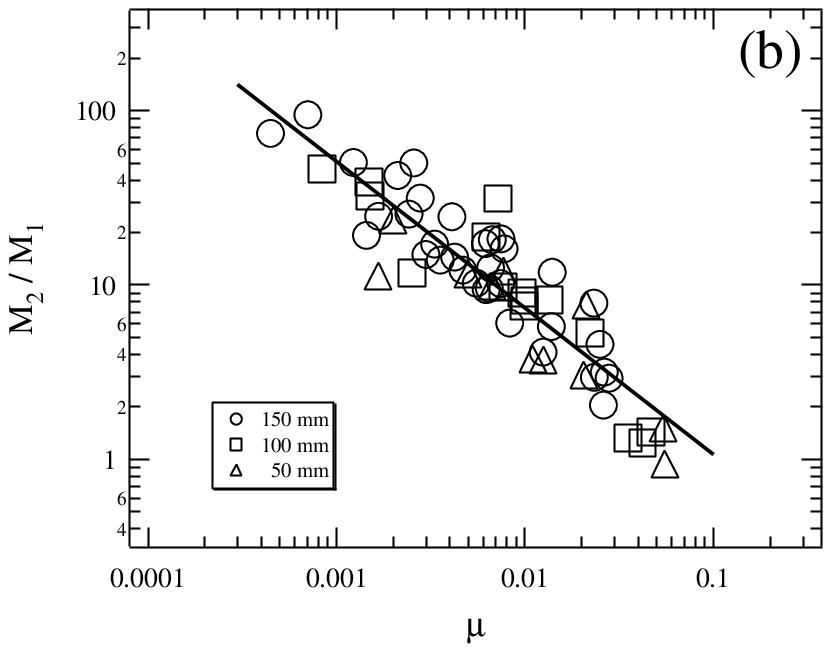}}
\caption{Critical behaviour of the weighted mean fragment mass $\frac{M_2}{M_1}$: (a) $\frac{M_2}{M_1}$ vs. $\epsilon$. (b) $\frac{M_2}{M_1}$ vs.\ $\mu$. The solid line indicates the form of the power-law $\frac{M_2}{M_1} \sim \mu^{-\sigma}$ ($\sigma = 0.84 \pm 0.05$).}
\label{fig:am}
\end{figure}

Since fragmentation events near the fragmentation critical point have less fragments, they cannot be analyzed statistically using only $N(m)$. Therefore, we used the $k$-th order moment of the fragment mass distribution, $M_k$, defined as
\begin{equation}
M_k = \sum_{m} m^{k}n(m).
\label{eq:mkdef}
\end{equation}
Since we are interested in fragmentation criticality, the height of the falling weight was controlled near the critical point. Despite our control of the height (imparted energy), the $M_k$ data fluctuate too much due to individual differences in the glass tube samples, e.g., the initial density of micro-cracks and residual stress distributions. When we considered the imparted energy as a control parameter, we could not find clear critical behavior quantitatively. For example, we show a log-log plot of  $\frac{M_2}{M_1}$ vs. $\epsilon$ in Fig.\ \ref{fig:am}(a). Where, $\frac{M_2}{M_1}$ is the weighted mean fragment mass. It is hard to say the data in Fig.\ \ref{fig:am}(a)  are particularly convincing about showing linear behavior. We could not consider that power-law fitting was appropriate with confidence. Alternatively, we used multiplicity $\mu$ analysis, which Campi introduced to the analysis of nuclei fragmentation \cite{Campi1}.  Multiplicity is defined as
\begin{equation}
\mu=m_{\min} \frac{M_0}{M_1},
\label{eq:mudef}
\end{equation}
where $M_0$, $M_1$, and $m_{\min}$ correspond to the number of fragments, the total mass of the fragments, and the smallest cut-off fragment mass, respectively. We fix $m_{\min}=0.01 \mbox{g}$ for all the analyses. This value usually corresponds to the smallest 2-dimensionality for the samples. According to this definition, the fragmentation critical point corresponds to $\mu = 0$. While the multiplicity is a resultant parameter, rather than a control parameter, it can be considered the control parameter that incorporates individual differences in the samples.  Furthermore, because the multiplicity is easy to measure and calculate, it has been applied to other fragmentation systems \cite{Campi1}. We use $\mu$ as a pseudo-control-parameter in this paper. Certainly, $\mu$ and the imparted energy are positively correlated, roughly. The larger the impact energy imparted, the larger $\mu$ becomes.

In Fig.\ \ref{fig:am}(b), we plot $\frac{M_2}{M_1}$ vs.\ $\mu$ of all the fragmentation event data.  It indicates the tendency toward divergence near the critical point.  The criticality appears independent of sample size. All the data match the same power-law line. Hence, we assume that the relationship between the weighted mean fragment mass $\frac{M_2}{M_1}$ and the multiplicity, $\mu$, holds as,
\begin{equation}
\frac{M_2}{M_1} \sim \mu^{- \sigma}.
\label{eq:sigmadef}
\end{equation}
The obtained value of $\sigma$ is $0.84 \pm 0.05$. This value is nontrivial because $\sigma \neq 1$.

\subsection{Scaling function}
\begin{figure}
\scalebox{1.0}[1.0]{\includegraphics{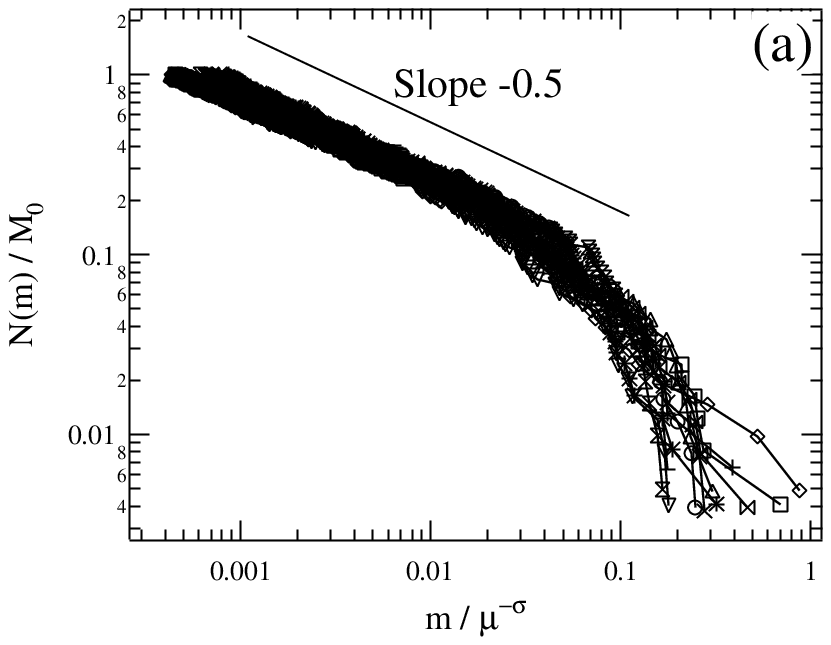}}
\scalebox{1.0}[1.0]{\includegraphics{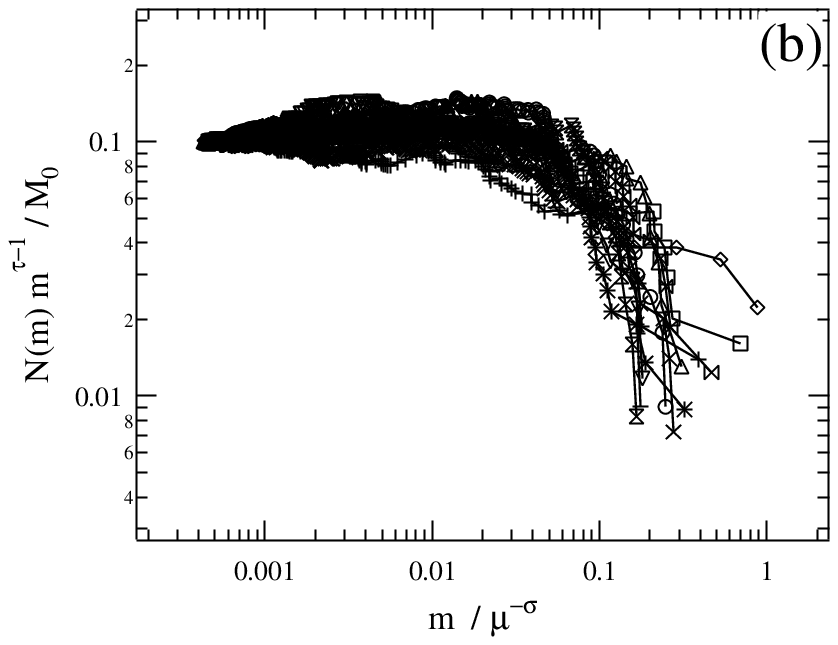}}
\caption{Plots of the scaling functions $f$ and $g$ approximated using (a) $N(m) / M_{0} \sim f( m / \mu^{-\sigma})$ and (b) $N(m) / M_{0} \sim m^{-(\tau - 1)} g( m / \mu^{-\sigma})$. The values of $\tau$ and $\sigma$ are obtained from scaling Figs.\ \ref{fig:cd} and \ref{fig:am}(b) as $1.5$ and $0.84$, respectively.}
\label{fig:sf}
\end{figure}

We discuss the possibility of scaling $N(m)$ curves using $\sigma$. For large $\mu$ data, $N(m)$ curves have the same power-law, in which the power is $\tau - 1 \simeq 0.5$. This value is independent of the size of the samples. This similarity allows us to collapse the $N(m)$ curves of different-sized samples into a unified scaling function. We expect the following scaling for distribution curves:
\begin{equation}
P(m) \sim f(m / \mu^{-\sigma}) \sim m^{-(\tau - 1)}g(m / \mu^{-\sigma}).
\label{eq:sf}
\end{equation}
Where $P(m)$ is the probability distribution of fragments whose masses are larger than $m$. We show the scaling result in Fig.\ \ref{fig:sf} with the approximation $P(m) = \frac{N(m)}{M_0}$. The curves in Fig.\ \ref{fig:sf} include the results for samples of three different sizes (11 curves of large $\mu$ data).  Figure \ref{fig:sf} shows good collapse of $N(m)$ curves into the scaling function $f$ and $g$. The function $f$ has a clear power-law regime, in which the power is $0.5$, and a unified cut-off scale. The former corresponds to the scaling region and the latter corresponds to exponential decay due to the finite size effect. Accordingly, the scaling function $g$ has a constant part and a rapidly decaying part. The flat part in $g$ exists only for the fully fragmented data, because clear power-law form $N(m) \sim m^{-(\tau-1)}$ can be satisfied with well fractured samples. The form of scaling function might change with $\mu$, however, we consider they will not depend on the sample size. We could not confirm the clear collapse for small $\mu$. More detailed experiments are necessary to verify the scaling functions in wide $\mu$ range. 

Bak et al. reported a similar scaling function for earthquake statistics \cite{Bak2}. While they discussed the time correlation of earthquakes, their scaling function is analogous to ours. They concluded that the unified scaling function for earthquakes depends only on the number of earthquakes occurring within the area and period considered. In Eq.\ \ref{eq:sf}, our scaling function depends on the variable $x= \frac{m} {\mu^{-\sigma}}$, which is the normalized mass of the fragments.  Moreover, the scaling Eq.\ \ref{eq:sf} resembles the scaling function {\AA}str\"om et al. obtained \cite{Astrom1}.

\subsection{Multi-scaling and simple biased cascade model}
\begin{figure}
\scalebox{1.0}[1.0]{\includegraphics{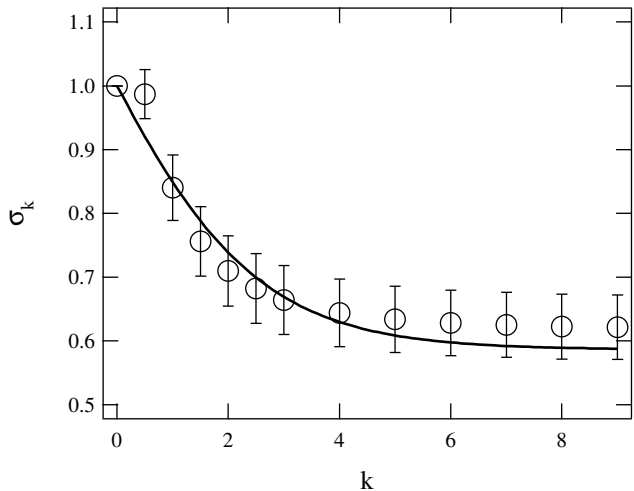}}
\caption{Higher-order weighted mean fragment mass scaling. The marks indicate the experimental results, and the solid curve indicates the result calculated using Eq.\ \ref{eq:biasedef} ($a=\frac{1}{3}$).}
\label{fig:sigma}
\end{figure}

Next, we discuss the behavior of the higher-order moments of the fragment mass distributions. We can easily expand Eq.\ \ref{eq:sigmadef} and introduce the higher-order weighted mean fragment mass scaling as,
\begin{equation}
\frac{M_{k+1}}{M_{k}} \sim \mu^{-\sigma_k}.
\label{eq:sigmakdef}
\end{equation}
The open circles in Fig.\ \ref{fig:sigma} show the $\sigma_k$ values obtained from fitting the all experimental data like Fig.\ \ref{fig:am}(b). While the value $\sigma_{k=0}$ is the trivial value $1$, $\sigma_{k\to\infty}$ approaches a nontrivial value of approximately $0.6$.

In order to understand this behavior, we consider a simple biased cascade model. Let us set a unit mass fragment initially. First, this unit mass is divided into two pieces of mass $a$ and $1-a$ $(0\leq a \leq 1)$. The same $a$-biased partition occurs for each fragment in the next step. This cascade continues until the imparted energy dissipates. From the above definition and Eq.\ \ref{eq:sigmakdef}, the following equation holds at each step:
\begin{equation}
\frac{a^{k+1} + (1-a)^{k+1}}{a^k + (1-a)^k} = 2^{-\sigma_k}.
\label{eq:biasedef}
\end{equation}
Substituting $a=\frac{1}{2}$ into Eq.\ \ref{eq:biasedef}, the trivial  value $\sigma_k =1$ is obtained for all $k$. This case corresponds to mono-scaling critical fragmentation. In Fig.\ \ref{fig:sigma}, however, $\sigma_k$ varies with $k$. This indicates a multi-scaling property of critical impact fragmentation. The solid curve in Fig.\ \ref{fig:sigma} is the value calculated using  Eq.\ \ref{eq:biasedef} with $a=\frac{1}{3}$. The curve and experimental data roughly agree within the error bars. Although this model is very simple, it explains the trend of the higher-order behavior. We conclude that fragmentation near the critical point has a multi-scaling nature. However, this only alters the problem. The origin of the symmetry breaking given with $a$ and the relation between $a$ (or $\sigma_k$) and $\tau$ remain unsolved. The cascade model does not seem to yield the appropriate value of $\tau$ ($\simeq 1.5$) directly. Since this model is too simple, it needs to be improved using 
more detailed analyses and experiments.

Similar multiplicative model was proposed for the energy cascade of eddy by Meneveau and Sreenivasan \cite{Meneveau1}. According to their result, the cascade of division into $0.3$ and $0.7$ can explain the energy cascade of turbulent flow. However, the origin of the value $0.3$ has remained unsolved. The value $0.3$ slightly agrees the value $\frac{1}{3}$ we obtained.

\section{Discussion}
Kun and Herrmann have analyzed the largest fragment mass as an order parameter \cite{Kun1}. In our experiments, the largest fragment mass had a very wide distribution, as seen in Fig.\ \ref{fig:sf}, while the weighted mean fragment mass behaved more calmly. Figure \ref{fig:cd} suggests that the power-law exponent might change with $\mu$. Ching et al. reported the impact energy dependence of the power-law exponent \cite{Ching1}. Their interest focused on the very large imparted energy region and not on the neighborhood of the fragmentation critical point. In addition, if we fit the power-law form to all $N(m)$ curves, the results involve large uncertainty, and do not show a clear trend. Ishii and Matsuhita reported that the distribution function for the low impact energy region has a log-normal form \cite{Ishii1}. We have not investigated the distribution of the concrete function form. The relationship between the two energy regions and the study of the concrete function form remain open to study.

Our experiments were focused on the 2-D fragmentation with the {\it flat} impact. Kun and Herrmann investigated the {\it point} impact 2-D fragmentation \cite{Kun1}. This might be a reason of the discrepancy between our results and their ones. Hayakawa suggested that the universality of the fragmentation depends not only on the dimensionality of fractured object, but also on the propagation dimensionality of the failure wave \cite{Hayakawa1}. {\AA}str\"om et al.\ performed numerical simulations of 2-D fragmentation with the {\it flat} expansion \cite{Astrom1}. Their results resemble ours in many points as described above. We believe that our results are valid for 2-D fragmentation by the {\it flat} impact to one side. Other dimensional objects or other failure wave propagation manner may produce other values of scaling exponents such as $\tau$ and $\sigma_k$. Moreover, scaling-law among these exponents may exist. As written in Sec.\ \ref{sec:introduction}, the exponent $\tau$ is independent of the material of fractured objects in fully fragmented state. We guess the other critical exponents are also independent of that, however, it should be examined. These problems are open questions. 

In general, the low impact energy region is difficult to study. Small differences of initial condition are enlarged by nonlinearity of fragmentation dynamics. Figure \ref{fig:am}(a) presents this difficulty well. Even in Fig.\ \ref{fig:am}(b), the fluctuating data can be seen. Some unrealized experimental parameters might not be controled. We think that the coincidence between the experimental result and the model shown in Fig.\ \ref{fig:sigma} is rough due to this reason.

In conclusion, we investigated the 2-D impact fragmentation of brittle solids experimentally. The measured characteristic exponents imply that the impact fragmentation of brittle solids does not belong to the percolation transition universality class. Instead, we found that the weighted mean fragment mass $\frac{M_2}{M_1}$ and the multiplicity parameter $\mu$ are related to the nontrivial scaling exponent $\sigma$. The cumulative fragment mass distributions of different-sized samples could be collapsed into the unified scaling function in the large $\mu$ regime. We calculated the generalized weighted mean fragment mass scaling $\frac{M_{k+1}}{M_k} \sim \mu^{-\sigma_{k}}$. The behavior of the multi-scaling exponent $\sigma_k$ was modeled using a simple biased cascade partition model. The critical behavior of other fragmentation systems should be studied in order to understand the universality of fragmentation phenomena in detail. Experiments with other materials, such as ceramics, 
or with samples of different dimensions may prove interesting.

We thank Professor H. Sakaguchi and Professor K. Nomura for their useful comments.

\end{document}